\documentclass[%
aip,
rsi,%
amsmath,
amssymb,
reprint,%
floatfix
]{revtex4-1}

\usepackage{graphicx}
\usepackage{dcolumn}
\usepackage{bm}

\usepackage{xcolor}

\begin{document}

\title[III-nitride on silicon
electrically injected microrings for nanophotonic circuits]{III-nitride on silicon
electrically injected microrings for nanophotonic circuits}

\author{Farsane Tabataba-Vakili}
\affiliation{Centre de Nanosciences et de Nanotechnologies, CNRS, Univ. Paris-Sud, Universit\'{e} Paris-Saclay, F-91120 Palaiseau, France}
\affiliation{Univ. Grenoble Alpes, CEA, INAC-Pheliqs, 38000 Grenoble, France}
\author{St\'{e}phanie Rennesson}
\affiliation{Universit\'{e} C\^{o}te d'Azur, CRHEA-CNRS, F-06560 Valbonne, France}
\author{Benjamin Damilano}
\affiliation{Universit\'{e} C\^{o}te d'Azur, CRHEA-CNRS, F-06560 Valbonne, France}
\author{Eric Frayssinet}
\affiliation{Universit\'{e} C\^{o}te d'Azur, CRHEA-CNRS, F-06560 Valbonne, France}
\author{Jean-Yves Duboz}
\affiliation{Universit\'{e} C\^{o}te d'Azur, CRHEA-CNRS, F-06560 Valbonne, France}
\author{Fabrice Semond}
\affiliation{Universit\'{e} C\^{o}te d'Azur, CRHEA-CNRS, F-06560 Valbonne, France}
\author{Iannis Roland}
\altaffiliation{Current address: Universit\'e Paris Diderot-Paris7, F-75013 Paris, France.}
\affiliation{Centre de Nanosciences et de Nanotechnologies, CNRS, Univ. Paris-Sud, Universit\'{e} Paris-Saclay, F-91120 Palaiseau, France}
\author{Bruno Paulillo}
\altaffiliation{Current address: ICFO-Institut de Ciencies Fotoniques, The Barcelona Institute of Science and Technology, 08860 Castelldefels, Spain}
\affiliation{Centre de Nanosciences et de Nanotechnologies, CNRS, Univ. Paris-Sud, Universit\'{e} Paris-Saclay, F-91120 Palaiseau, France}
\author{Raffaele Colombelli}
\affiliation{Centre de Nanosciences et de Nanotechnologies, CNRS, Univ. Paris-Sud, Universit\'{e} Paris-Saclay, F-91120 Palaiseau, France}

\author{Moustafa El Kurdi}
\affiliation{Centre de Nanosciences et de Nanotechnologies, CNRS, Univ. Paris-Sud, Universit\'{e} Paris-Saclay, F-91120 Palaiseau, France}
\author{Xavier Checoury}
\affiliation{Centre de Nanosciences et de Nanotechnologies, CNRS, Univ. Paris-Sud, Universit\'{e} Paris-Saclay, F-91120 Palaiseau, France}
\author{S\'ebastien Sauvage}
\affiliation{Centre de Nanosciences et de Nanotechnologies, CNRS, Univ. Paris-Sud, Universit\'{e} Paris-Saclay, F-91120 Palaiseau, France}
\author{Laetitia Doyennette}
\affiliation{Laboratoire Charles Coulomb (L2C), Universit\'e de Montpellier, CNRS, Montpellier, France}
\author{Christelle Brimont}
\affiliation{Laboratoire Charles Coulomb (L2C), Universit\'e de Montpellier, CNRS, Montpellier, France}
\author{Thierry Guillet}
\affiliation{Laboratoire Charles Coulomb (L2C), Universit\'e de Montpellier, CNRS, Montpellier, France}
\author{Bruno Gayral}
\affiliation{Univ. Grenoble Alpes, CEA, INAC-Pheliqs, 38000 Grenoble, France}
\author{Philippe Boucaud}
\email{philippe.boucaud@crhea.cnrs.fr}
\affiliation{Universit\'{e} C\^{o}te d'Azur, CRHEA-CNRS, F-06560 Valbonne, France}

\begin{abstract}
Nanophotonic circuits using group III-nitrides on silicon are still lacking one key component: efficient electrical injection. In this paper we demonstrate an electrical injection scheme using a metal microbridge contact in thin III-nitride on silicon mushroom-type microrings that is compatible with integrated nanophotonic circuits with the goal of achieving electrically injected lasing. Using a central buried n-contact to bypass the insulating buffer layers, we are able to underetch the microring, which is essential for maintaining vertical confinement in a thin disk. We demonstrate direct current room-temperature electroluminescence with $440\text{~mW/cm}^2$ output power density at 20 mA from such microrings with diameters of 30 to 50 $\mu$m. The first steps towards achieving an integrated photonic circuit are demonstrated.
\end{abstract}

\maketitle

\section{Introduction}

The group III-nitride material system can emit light from the deep ultraviolet (UV) to visible (VIS) spectral range \cite{Strite1992}, opening an extraordinary realm of possible applications for active photonic devices. Passive devices from the UV to the infra-red (IR) are conceivable thanks to the very large transparency window of III-nitrides. Opto-electronics based on III-nitrides, including blue/violet light emitting diodes (LEDs) and laser diodes (LDs) and white light LEDs are already being mass produced for consumer applications, such as general lighting and multi-color displays \cite{Nakamura2013}. Potential applications for III-nitride nanophotonics, that is photonics utilizing microcavities, waveguides, and other devices interacting with light on the sub-wavelength scale,  range from quantum technologies, such as single photon sources \cite{Holmes2014}, ion trapping \cite{Lucas2004}, and parametric down conversion \cite{Javurek2014} to bio-sensing, III-nitrides being bio-compatible \cite{Steinhoff2003, Hofstetter2012}.  Over the past 15 years, there have been numerous demonstrations of III-nitride nanophotonic devices, including high quality factor optical microcavities \cite{Simeonov2008, Mexis2011, Trivino2014, Rousseau2018}, microlasers \cite{Kneissl2004, Choi2006, Tamboli2007, Simeonov2007, Athanasiou2014, Zhang2015, Trivino2015, Selles2016, Selles20162, Athanasiou2017, Feng2018}, and both passive \cite{Pernice20122, Jung2013, Stegmaier2014, Bruch2015, Lu2018, Liu2018} and active \cite{Shi2017, Gao2017, TabatabaVakili2018} photonic circuits on sapphire, silicon carbide, or silicon substrates.

Electrical injection was utilized only in very few cases in III-nitride microdisk cavities or photonic circuits \cite{Kneissl2004, Shi2017, Gao2017, Feng2018}, yet a main interest in using III-nitrides is to have active emitters, which are much more feasible for applications if electrically driven as opposed to optically pumped. Efficient electrical injection continues to pose a challenge in thin III-nitride layers. However, the advantage of using silicon as a substrate for III-nitride nanophotonics is the possibility of monolithic integration into a well-established photonic platform, extending it to encompass UV-VIS emitters. 
 Major difficulties in achieving low threshold and long lifetime electrically injected lasing are to sufficiently reduce the threading dislocation density (TDD) (usually in the order of $10^{9}-10^{10}~\text{cm}^{-2}$) in gallium nitride (GaN) grown on silicon which is caused by the large lattice mismatch of 17\% and to reduce cracking due to the large difference in thermal expansion coefficient of 116\% between GaN and silicon \cite{Sun2016, Jeon2015}.  A significant dislocation density does not prevent to observe lasing as shown by Sun et al. \cite{Sun2016} who in 2016 demonstrated the first ever laser diode in the III-nitride on silicon material system with a TDD of $6 \cdot 10^8 \text{~cm}^{-2}$. Lasing has also been achieved with higher TDDs \cite{Nakamura1998}. A standard GaN on silicon epilayer can thus be an appropriate test platform for nanophotonic circuits incorporating active laser devices, even if it has limitations for reliability under continuous wave electrical injection.

The field of III-nitride on silicon nanophotonics will greatly benefit from an  electrical injection scheme compatible with thin epilayers. There have been some attempts at electrical injection in III-nitride photonic circuits for light communication \cite{Shi2017, Gao2017}, however these have been on impractically large devices. Feng et al. \cite{Feng2018} recently demonstrated lasing under electrical injection in III-nitride "sandwich"-type microdisks on silicon, however, they are using $5.8~\mu\text{m}$ thick heterostructures and side contacts that do not allow for fabrication of high quality, efficient nanophotonic circuits. The need for side contacts can be avoided by via etching, which was demonstrated by Hikita et al. for field effect transistors using a lateral device structure \cite{Hikita2005, Hikita2007}. A backside substrate and buffer layer removal approach for all vertical power diodes was proposed by Zhang et al. \cite{Zhang2018}. 

\section{Fabrication}

\begin{figure}[ht!]
\centering\includegraphics[width=1\linewidth]{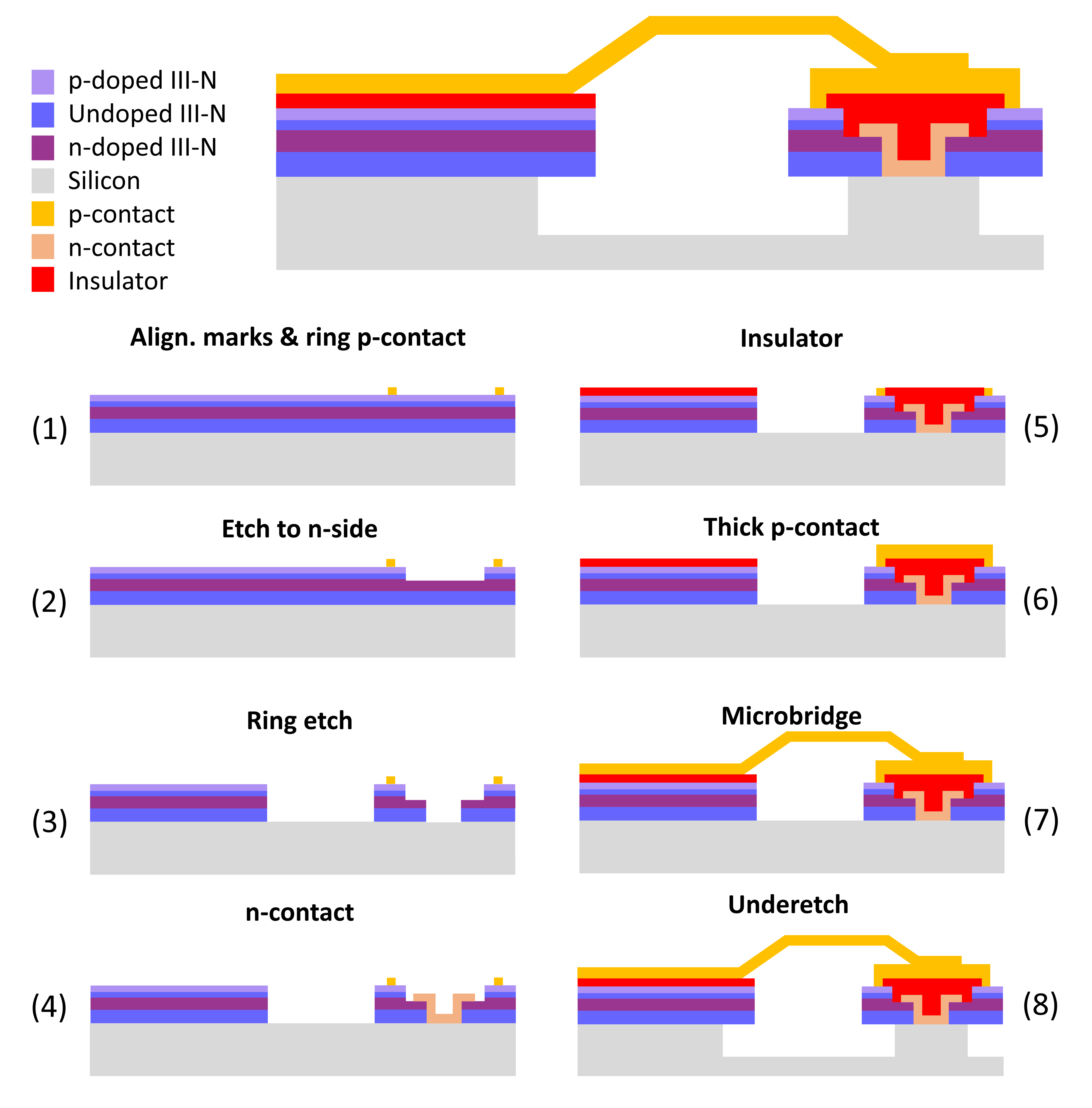}
\caption{Process flow of microrings for electrical injection.}
\label{fig:process}
\end{figure}

In this paper, we are proposing a vertical device with an electrical injection scheme that is compatible with thin layers ($1-2~\mu\text{m}$) and allows for the fabrication of side-coupled bus waveguides with small gaps (less than 100 nm) necessary for efficient coupling in the blue to UV spectral range. We use mushroom-type microrings with a top buried n-contact that bypasses the insulating buffer layers currently required in our epitaxial scheme for growth of high material quality GaN on silicon. A mechanically stable metallic microbridge contact allows for easy probing of a larger p-pad. The back contact is taken from the n-doped silicon, without metalization. This technology is scalable to complex integrated photonic circuits and will provide an asset to the field.

\begin{figure*}[ht!]
\centering\includegraphics[width=0.8\linewidth]{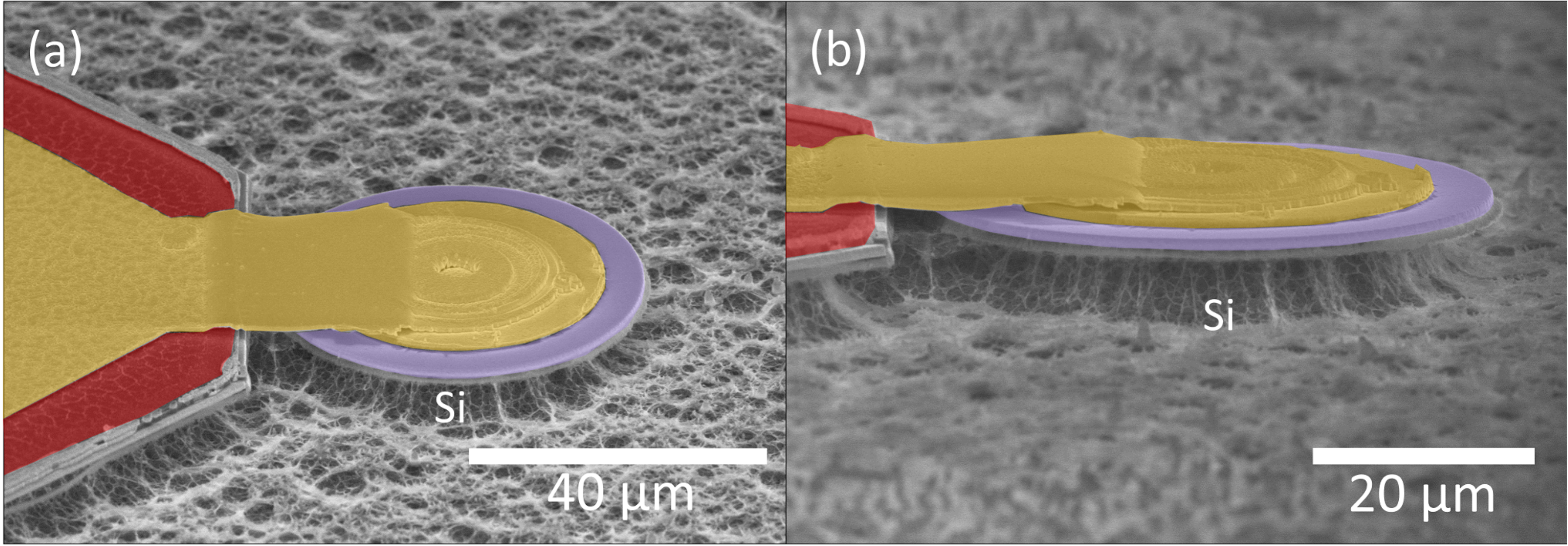}
\caption{a) False color SEM images of a  $50~\mu\text{m}$ diameter microring for electrical injection. b) Same device but from a different angle to emphasize the underetching of the microring. The metal microbridge and p-contact are highlighted in yellow, the insulator in red, and the III-nitride in purple. The rough gray area is the etched silicon.}
\label{fig:sem}
\end{figure*}

Microrings with $30$ to $50~\mu\text{m}$ diameter were fabricated using an 8-step optical and e-beam lithography process depicted schematically in Fig. \ref{fig:process}. The dimensions of different layers were varied throughout the mask. First (step 1) alignment marks and 2 to $5~\mu\text{m}$ wide ring Ni/Au p-contacts were deposited and annealed at $500^\circ\text{C}$ for 5 min. Next (step 2) the center of the p-ring is etched to the n-side using inductively coupled plasma (ICP) etching with chlorine ($\text{Cl}_2$) and boron trichloride ($\text{BCl}_3$) gases. Using e-beam lithography with UVIII resist (step 3), the microring is defined leaving 2 to $6~\mu \text{m}$ free between the ring p-contact and the edge of the disk. The pattern is subsequently etched into a plasma-enhanced chemical vapor deposited (PECVD) silicon dioxide ($\text{SiO}_2$) hard mask using reactive ion etching (RIE) and into the III-nitride and down to the silicon substrate using ICP etching. Next (step 4) a circular Ti/Al/Ni/Au n-contact is deposited at the center of the ring, connecting the n-GaN to the silicon substrate, bypassing the insulating buffer layers. The distance between the n-contact and the edge of the microring is between 7 and $16~\mu\text{m}$. Subsequently, (step 5) $\text{SiO}_2$ is deposited as an insulating layer and etched away at the edge of the disk and on the p-ring using RIE. A thick circular Ni/Au p-contact is deposited over the center of the disk, covering the ring p-contact and the insulator (step 6). Next (step 7) using a 2-step lithography involving a photoresist temperature gradient reflow step to form the bridge holder, the microbridge contact and p-pad are defined and a thick Ti/Au layer is deposited. As a final step (step 8) the microrings are underetched using xenon difluoride ($\text{XeF}_2$) gas. A further reduction in device size could be realized by switching to a full e-beam lithography process. Devices in the $10~\mu \text{m}$ diameter range would be feasible. Finding a way to make the entire heterostructure conductive can also decrease device size, as it would allow for the fabrication of standard mushroom-type microdisks \cite{Mexis2011}.

False color scanning electron microscopy (SEM) images of a tilted $50~\mu\text{m}$ diameter microring are shown in Fig. \ref{fig:sem} for two different angles. The undercut of the microring and the bridge suspension are clearly visible. The rough area around the ring is the silicon substrate after isotropic etching by $\text{XeF}_2$ gas. 

\section{Experiment and discussion}

\begin{figure*}[ht!]
\centering\includegraphics[width=0.8\linewidth]{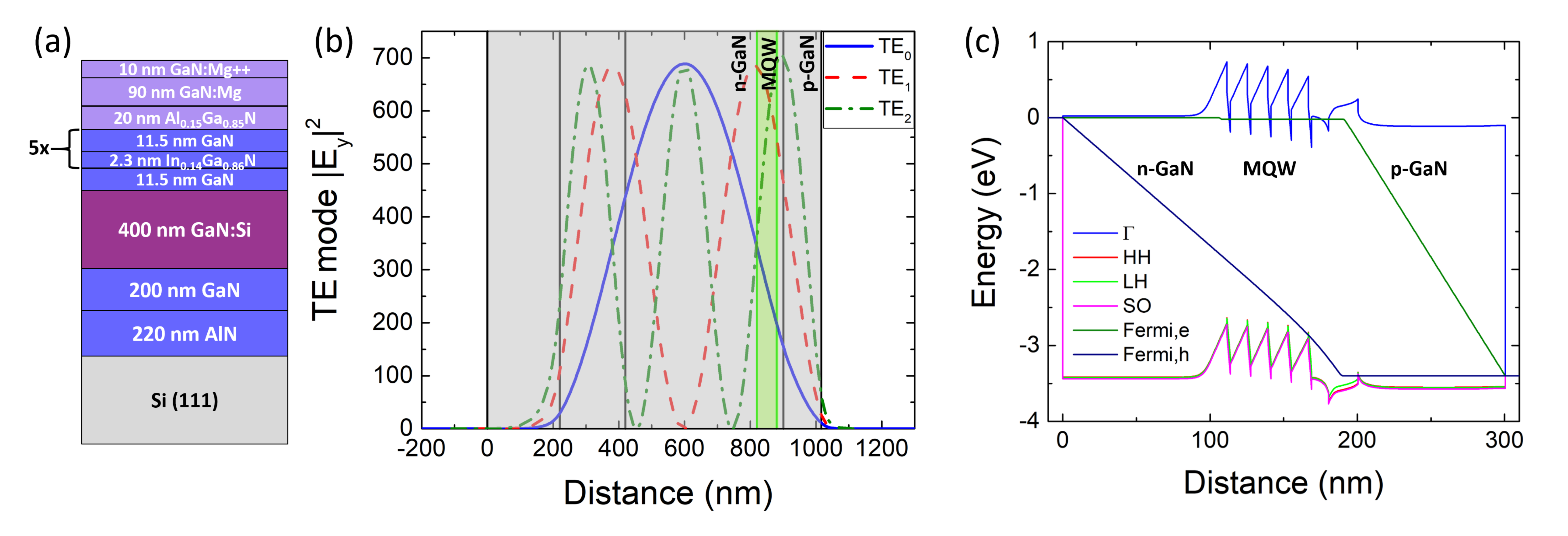}
\caption{The sample structure: a) Detailed heterostructure, b) Simulated mode confinement of the TE$_0$, TE$_1$, and TE$_2$ modes, the MQW region is highlighted in green, the heterostructure in gray, c) Energy band structure of the sample at 3.4 V bias.}
\label{fig:structure}
\end{figure*}

The investigated sample was grown on a 2 inch n-type (doping concentration $5\cdot 10^{18} \text{cm}^{-3}$)  silicon (111) wafer using metal organic chemical vapor deposition (MOCVD). The heterostructure is illustrated in Fig. \ref{fig:structure} (a). First an undoped buffer layer of 220 nm aluminum nitride (AlN) and 200 nm GaN was grown, followed by the LED heterostructure consisting of 400 nm GaN:Si (doping level $5\cdot 10^{18}~ \text{cm}^{-3}$), 11.5 nm of undoped GaN, 5 pairs of 2.3 nm $\text{In}_{0.14}\text{Ga}_{0.86}\text{N}$ quantum wells (QWs) and 11.5 nm GaN barriers, a 20 nm Mg-doped $\text{Al}_{0.15}\text{Ga}_{0.85}\text{N}$ electron blocking layer (EBL) (doping concentration 
$4\cdot 10^{19} ~\text{cm}^{-3}$), 90 nm of GaN:Mg (doping concentration 
$4\cdot 10^{19} ~\text{cm}^{-3}$), and 15 nm of GaN:Mg (doping concentration 
$1\cdot 10^{20} ~\text{cm}^{-3}$).  The TDD estimated from X-ray diffraction (XRD) and atomic force microscopy (AFM) is in the range of  $5 \cdot 10^{9}$ to $2 \cdot 10^{10}~\text{cm}^{-2}$ and the root mean square (rms) roughness is 1 nm as determined by an atomic force microscopy (AFM) scan of $10~\mu \text{m} \times 10~ \mu \text{m}$. The device scheme proposed in this paper allows for much simpler heterostructures than required for the devices reported on by Feng et al. \cite{Feng2018}, as no cladding layers are used.

Simulation of the vertical mode confinement of the TE$_0$, TE$_1$, and TE$_2$ modes are shown in Fig. \ref{fig:structure} (b) for an underetched disk. The overlap of these modes with the QWs (without barriers) are 0.96 \%, 2.4 \%, and 2.0 \%, respectively. By increasing the thickness of the p-GaN cap layer, for example, we could increase this overlap of the TE$_0$ mode to a maximum of 1.6 \% for a 400 nm cap thickness. Laterally the mode is confined by total internal reflection at the edge of the disk, extending over a few microns. A simulation of the band structure is shown in Fig. \ref{fig:structure} (c) under a 3.4 V bias, which corresponds to the GaN bandgap. Simulations of the radiative recombination show that the carrier density is higher in the QWs closer to the p-side and thus fewer wells may lead to an enhanced performance \cite{David2008}.

\begin{figure}[ht!]
\centering\includegraphics[width=0.8\linewidth]{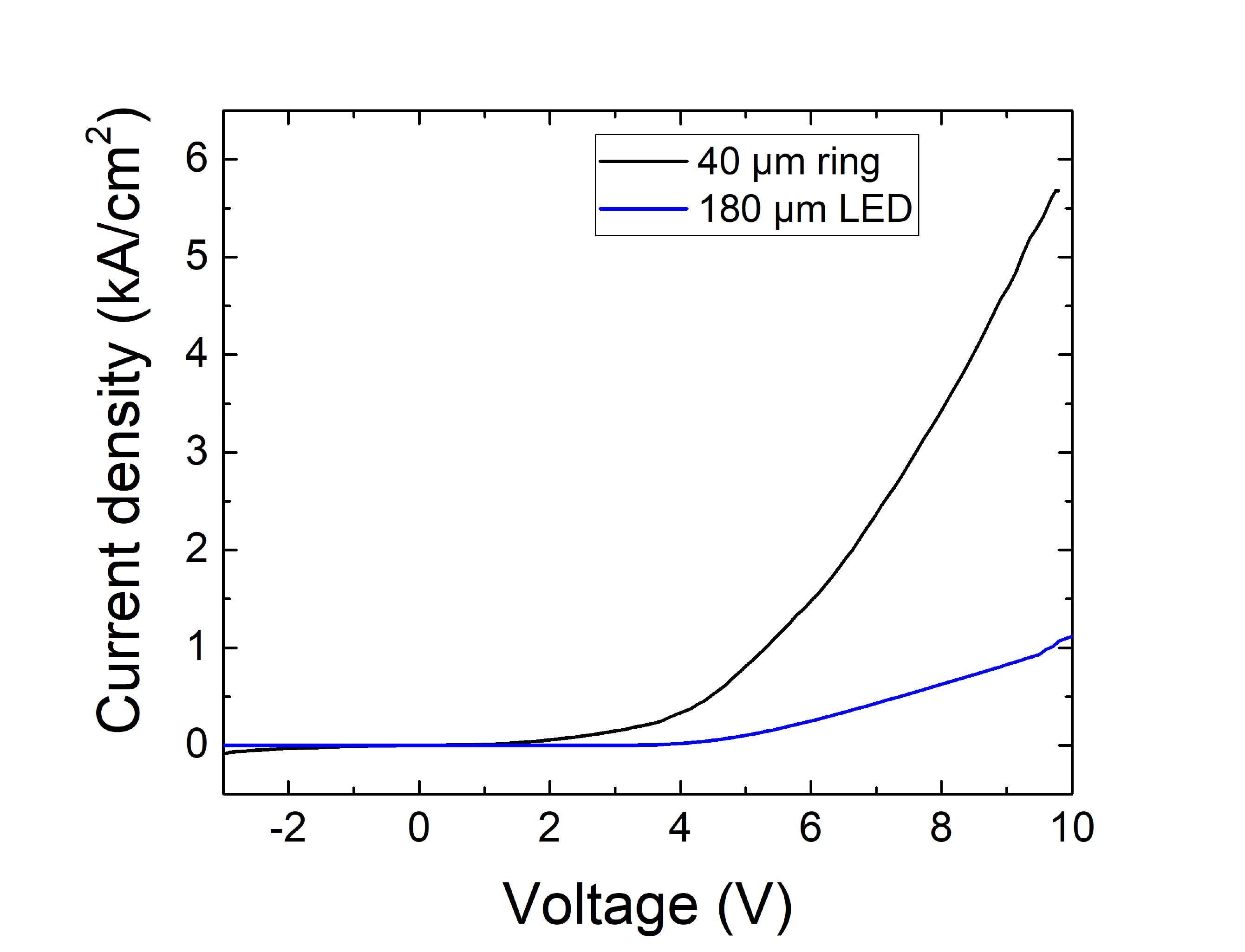}
\caption{JV curves of a $40~\mu\text{m}$ diameter microring and a $180~\mu\text{m}$ lateral LED for reference.}
\label{fig:IV}
\end{figure}

A typical current density over voltage (JV) curve of a $40~\mu\text{m}$ diameter microring is shown in Fig. \ref{fig:IV} and is compared to the JV curve of a lateral LED with a $180~\mu\text{m}$ diameter p-contact and top n-contact on the n-GaN. The turn-on voltage is similar in both cases, which shows that the microring process with its multiple etch steps and small device dimensions, as well as taking the n-contact through the silicon substrate, do not have any negative influence on the device performance. The surface area of both p-contacts is very different, and a direct comparison of the current densities at a given voltage is not straightforward. For microrings the current density quickly becomes elevated. The overall high voltage in forward direction is due to the resistive p-GaN. The microring also shows a slightly leaky behavior in reverse bias. It is important to note that for the as-grown sample and when fabricating large mesas of 1 mm diameter, vertical conductivity through the buffer layers is observed even if the buffer is expected to be insulating, while mesas of around $200~\mu\text{m}$ diameter are vertically insulating. This indicates that the vertical conductivity is not simply driven by the dislocation density. We attribute this phenomenon to current flowing through macroscopic defects in the buffer layer and at the AlN/Si interface that might result from dislocation formation, dislocation bundles, inter-diffusion, and possibly cracks in large mesas. It also shows that the approach with the central annular contact is mandatory for micrometer-size devices. Meanwhile the central annular contact connected to the silicon substrate allows current injection through the substrate, which is an important technical option for III-nitride on silicon devices.

\begin{figure*}[ht!]
\centering\includegraphics[width=0.8\linewidth]{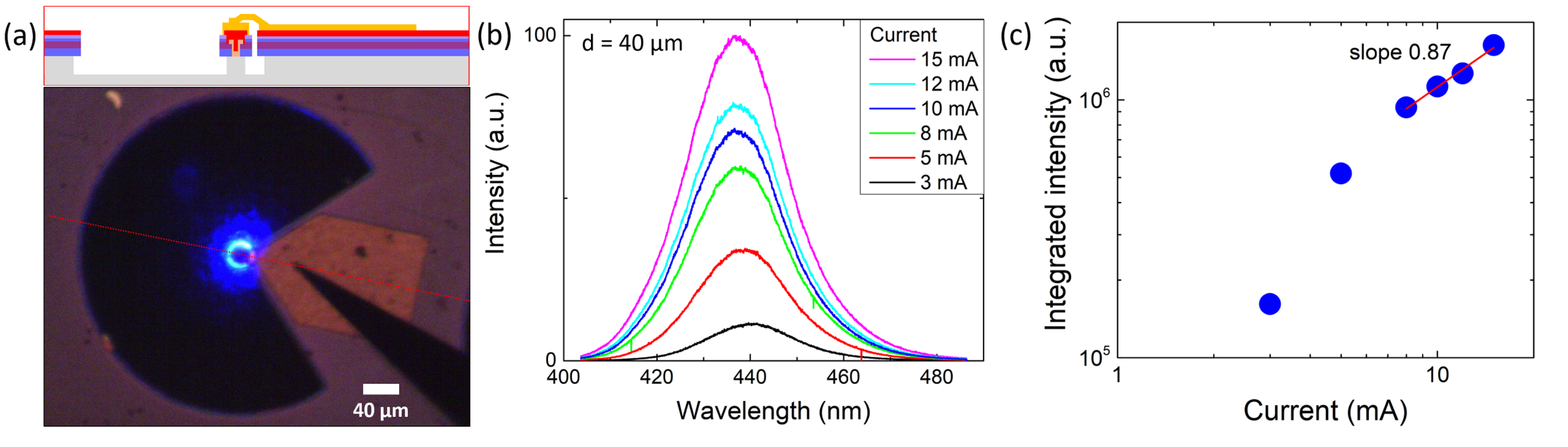}
\caption{Measurement results: a) Top: Cross-sectional view of the device along the red dashed line. Bottom: Photo of a powered $40~\mu\text{m}$ diameter device emitting blue electroluminescence. b) Spectra of a device with a $40~\mu\text{m}$ diameter at different injection currents. Maximum current density $4.2~\text{kA/cm}^2$. c) Integrated intensity of the measurements in b).}
\label{fig:EL}
\end{figure*}

We performed direct current room-temperature electroluminescence (EL) measurements on these devices, using probe tips placed on the large p-pad and near the sample on the copper block for the n-contact through the silicon substrate. The emission is collected from the side of the sample with a microscope objective and the light is guided through air into a grating monochromator with a liquid nitrogen cooled charge-coupled device (CCD) as detector. Figure \ref{fig:EL} (a) shows a photo of a $40~\mu\text{m}$ microring emitting blue EL. A cross-sectional view of the device along the red dashed line is shown at the top of the figure. The microring is at the center of a black 3/4 circle, which is etched to the Si substrate by ICP and isotropically etched by the $\text{XeF}_2$ gas. On the right side of this circle is a 140 x 140 $\mu\text{m}^2$ p-pad that connects to the microring via a microbridge. Figure \ref{fig:EL} (b) shows spectra of a $40~\mu\text{m}$ diameter ring for currents from 3 to 15 mA measured using a 600 grooves/mm grating, an $80~\mu\text{m}$ slit width, and 0.01 s integration time. A slight blue shift is observed with increasing current. The maximum current of $15~\text{mA}$ corresponds to a current density of $4.2 ~\text{kA/cm}^2$, which is not sufficient to achieve lasing. Device degradation and breakdown are observed at currents between 15 and 30 mA, likely due to the not optimized insulator between the p- and n-contacts at the center of the microdisk. By using silicon nitride (SiN) instead of SiO$_2$ better device performance should be attained. Feng et al. reported a lasing threshold of $25~\text{kA/cm}^2$~for devices with $40~\mu\text{m}$ diameter, while Kneissl et al. observed lasing at $7.8 ~\text{kA/cm}^2$ for $500~\mu\text{m}$ diameter devices on sapphire \cite{Feng2018, Kneissl2004}. Under optical pumping we have observed rather high thresholds of around $3 ~\text{mJ/cm}^2$ per pulse for $5~\mu\text{m}$ diameter disks \cite{Selles20162} fabricated on similar material. No whispering gallery modes (WGMs) are observed using a 3600 grooves/mm grating with a spectral resolution of 0.02 nm. The free spectral range (FSR) of first-order modes in such a $40~\mu\text{m}$ diameter ring is expected to be around 0.4 nm, as extrapolated from finite-difference time-domain (FDTD) simulations for smaller devices. However, there are many competing families of modes, and it is not evident that the large spectral density of modes allows for an observation of WGMs below threshold. Both Kneissl et al. and Feng et al., who used devices in a similar diameter range do not observe any modes below threshold \cite{Kneissl2004,Feng2018}. Hole diffusion is an issue in such devices due to the distance from the p-ring to the edge of the microdisk and tunnel junctions will be investigated as a potential solution. Figure  \ref{fig:EL} (c) shows a double logarithmic plot of the integrated intensity over current, depicting a slope of around 1 at high current injection, suggesting dominant radiative recombination, while non-radiative processes dominate at low current. This implies a good material quality, which could be further improved by combining a 3D growth mode during the buffer layer growth to reduce the TDD with growth on mesa patterned templates to reduce the tensile stress in the GaN layers to prevent their cracking
\cite{Hossain2012, Tanaka2017}.  We performed polarization measurements (not shown) at 0, 45, and $90^\circ$, indicating that the microring emission is mainly of transverse electric (TE) polarization, as expected for InGaN/GaN QWs \cite{Frankowsky1996}.  Using an integrating sphere we determined the output power of a $40~\mu\text{m}$ diameter microring corresponding to the IV response in Fig. \ref{fig:IV} to be $2~\mu\text{W}$ at 20 mA, giving an output power density of $440\text{~mW/cm}^2$ at a current density of $5.7\text{~kA/cm}^2$. This power density is below the one reported for microLEDs grown on sapphire \cite{Gossler2014}, but in the latter case a full optimization of collection efficiency is performed. Here the microring radiates preferentially in the layer plane, which is detrimental for collection efficiency, but an asset for in-plane coupling and nanophotonic circuits, as described below. 

\section{Perspectives}

\begin{figure}[ht!]
\centering\includegraphics[width=0.8\linewidth]{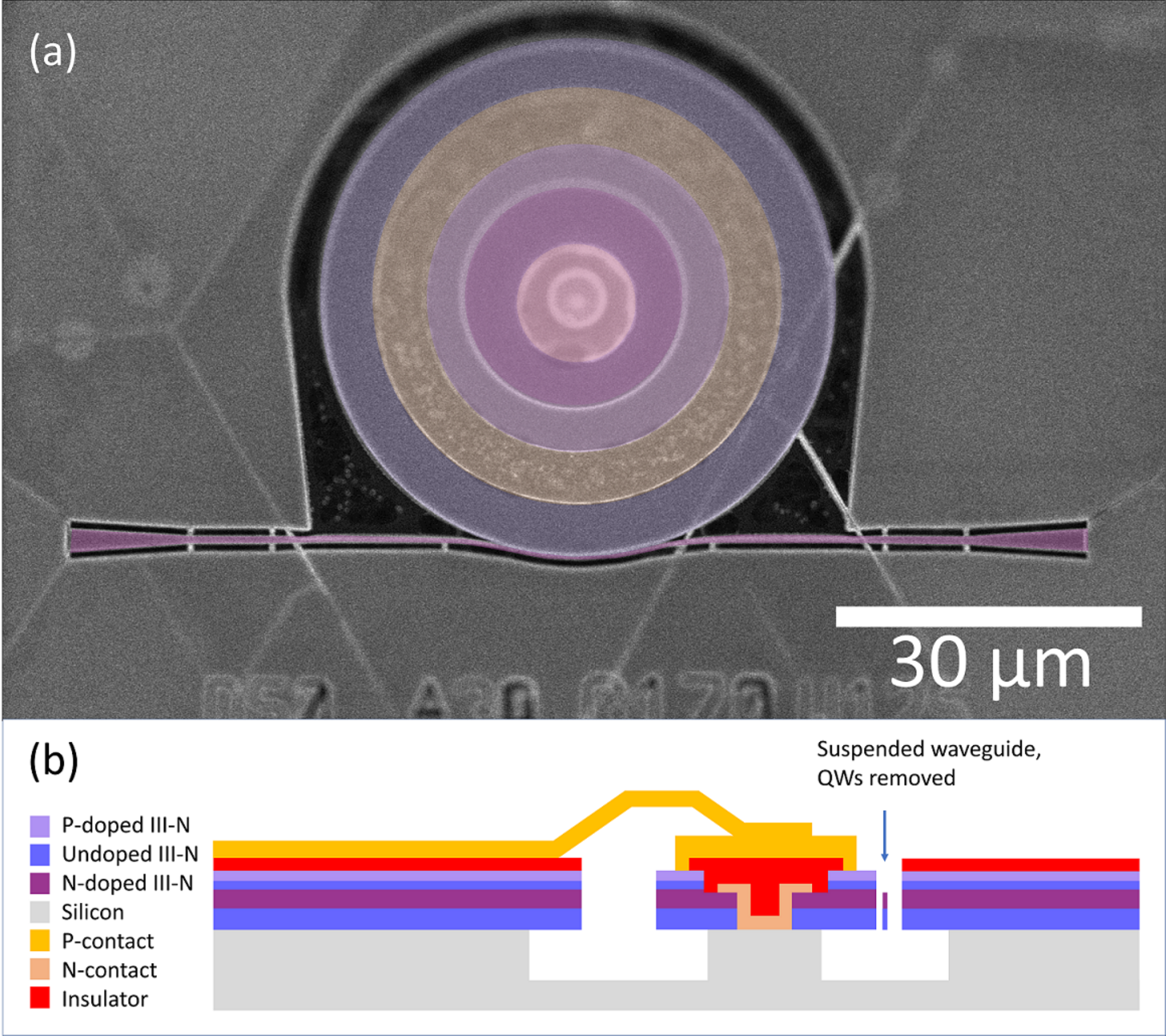}
\caption{Microring under electrical injection with bus waveguide side coupling: a) False color SEM image of device with a 50 $\mu$m diameter ring and a 100 $\mu$m long bus waveguide after several processing steps. b) Side-view sketch of a full device.}
\label{fig:ringelec}
\end{figure}

In order to achieve lasing in such microring structures next steps will include a) replacing the $\text{SiO}_2$ insulating layer with SiN to delay electrical breakdown, b) improving the lateral current spreading through the use of tunnel junctions, c) improving the vertical confinement and mode overlap by replacing the n-GaN layer with n-AlGaN, and d) lowering the dislocation density through selective area growth.

Furthermore, we envision to demonstrate a photonic circuit consisting of an electrically injected microring, a side-coupled bus waveguide with a small gap in the order of 100 nm, and grating out-couplers terminating the waveguide, as illustrated in Fig. \ref{fig:ringelec}, and as previously demonstrated under optical pumping \cite{TabatabaVakili2018}. This process is very challenging, as alignment precision between 20 nm for the e-beam lithography steps and $1~\mu \text{m}$ for the optical lithography steps are required, and proximity effect correction is needed when defining the microring and bus waveguide. Note that the etching of small gaps and narrow waveguides, which is essential for efficient evanescent coupling in the blue, requires the total heterostructure thickness to be less than $2~\mu \text{m}$. Such devices will allow to demonstrate the viability of the III-nitride on silicon nanophotonic platform for real-world applications.

\section{Conclusion}

In conclusion, we have demonstrated a scheme for electrical injection in microrings in thin III-nitride epilayers on silicon that is compatible with microlaser diodes coupled to integrated photonic circuits as it allows for device underetching and the possibility of fabricating side-waveguides with small gaps of less than 100 nm, necessary for efficient coupling in the blue-UV spectral range. We have demonstrated EL of such microrings contacted with a metallic microbridge with $440\text{~mW/cm}^2$ output power density at 20 mA.

\section*{Acknowledgments}
We acknowledge the support of W.Y. Fu, K.H. Li, Y.F. Cheung, and H.W. Choi at Hong Kong University (HKU) with the power measurement with an integrating sphere. This work was supported by Agence Nationale de la Recherche
under the MILAGAN convention (ANR-17-CE08-0043-02).
This work was also partly supported by the RENATECH
network. We acknowledge support by a public grant overseen
by the French National Research Agency (ANR) as part of the
"Investissements d'Avenir" program: Labex GANEX (Grant
No. ANR-11-LABX-0014) and Labex NanoSaclay (reference:
ANR-10-LABX-0035).



\bibliography{mybib}

\end{document}